\documentclass[aps,nofootinbib,prd,twocolumn,preprintnumbers]{revtex4-1}
\pdfoutput=0 

\usepackage[T1]{fontenc} 
\usepackage{graphicx}
\usepackage{epsfig}
\usepackage{rotating}
\usepackage{color}
\usepackage[normalem]{ulem}
\usepackage{multirow}
\usepackage{tabularx}
\usepackage{graphicx}
\usepackage{dcolumn}
\usepackage{hyperref}
\usepackage{float}
\usepackage{amsmath}
\usepackage{amssymb}
\usepackage{ulem}
\usepackage{color}

\begin{document}

\title{\bf Impact of Non-standard Interactions on Neutrino-Nucleon Scattering}

\author{D.K. Papoulias~$^{1}$}\email{dimpap@cc.uoi.gr}
\author{T.S. Kosmas~$^1$}\email{hkosmas@uoi.gr}

\affiliation{$^1$~Theoretical Physics Section, University of Ioannina, GR-45110 Ioannina, Greece}

\begin{abstract}
Non-standard neutrino-nucleon interaction is formulated and explored within the energy range of quasi-elastic scattering. In
particular, the study focuses on the neutral-current elastic (anti)neutrino scattering off nucleons described by the exotic reactions
$\nu_\alpha ({\bar \nu}_\alpha) + n \rightarrow  \nu_\beta ({\bar \nu}_\beta) + n $ and $ \nu_\alpha ({\bar \nu}_\alpha) +  p  \rightarrow  \nu_\beta  ({\bar \nu}_\beta) + p$, 
which provide corrections to the dominant Standard Model processes.
In this context, it is shown that the required exotic nucleon form factors may have a significant impact on the relevant cross sections.
Besides cross sections, the event rate is expected to be rather sensitive to the magnitude of the lepton-flavour violating parameters
resulting in an excess of events. The overlap of non-standard interactions and strange quark contributions, in the region of few GeV neutrino energies, is also examined. The formalism is applied for the case of the relevant neutrino-nucleon scattering experiments (LSND, MiniBooNE, etc.) and motivates the notion that such facilities have high potential to probe NSI.
\end{abstract}
\maketitle


\section{Introduction} 

Neutrinos are among the most elusive particles in nature  
and in order to investigate their properties~\cite{Alberico:2001sd,Morfin:2012kn} various terrestrial detectors have been built~\cite{Abe:2010hy}.  These ghostly 
particles fill the whole universe and reach Earth coming 
from the sun (solar neutrinos), from supernova explosions (supernova
neutrinos) and from many other celestial objects (e.g. black hole binary
stars, active galactic nuclei, etc.)~\cite{Smponias:2015nua}. The majority of them pass through Earth and through the nuclear detectors, designed for such purposes, 
without leaving any trace or signal~\cite{Balasi:2015dba}. This is mostly due to the fact that neutrinos interact 
very weakly with matter~\cite{Valle:2015pba,Kosmas:1996fh}. For the investigation of neutrino-matter scattering, it is feasible to employ powerful accelerators operating at major laboratories such as Fermilab, J-Park, CERN, etc. These facilities can produce intensive neutrino beams of which a tiny fraction can be detected by novel detectors placed in the beam line (e.g. COHERENT experiment at Oak
Ridge~\cite{Akimov:2015nza}, TEXONO experiment in Taiwan~\cite{Kerman:2016jqp}, $\nu$GeN~\cite{Belov:2015ufh} and GEMMA~\cite{Beda:2013mta} experiments in Russia and CONNIE project in Brazil~\cite{Moroni:2014wia,Aguilar-Arevalo:2016qen}).

On the theoretical side, phenomenological models within and beyond the standard electroweak theory come out with theoretical predictions for many aspects of neutrinos in trying to understand their properties and interactions~\cite{Schechter:1981hw}, and propose appropriate neutrino probes for extracting new experimental results~\cite{Kosmas:1989pj,Kosmas:2015sqa,Kosmas:2015vsa}. Current important areas of research concern the neutrino masses~\cite{Boucenna:2014zba}, neutrino oscillations~\cite{Tortola:2012te,Forero:2014bxa}, neutrino electromagnetic properties~\cite{Shrock:1982sc,kayser:1982br,Nieves:1981zt,Beacom:1999wx,Broggini:2012df}, and so forth, their role in the evolution of astrophysical sources such as the sun or supernovae~\cite{Balasi:2011zz,Tsakstara:2011zzc,Ydrefors:2012hh}, their impact on cosmology (e.g. in answering the question of the matter-antimatter asymmetry of the universe) and others.

%
\begin{figure*}[t]
\centering
\includegraphics[width=0.49 \textwidth]{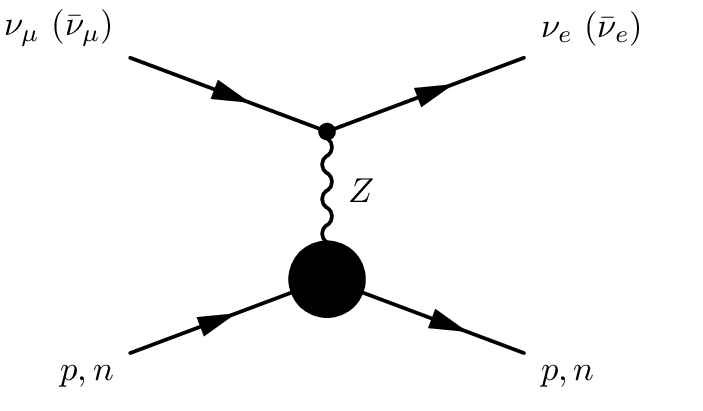}
\includegraphics[width=0.49 \textwidth]{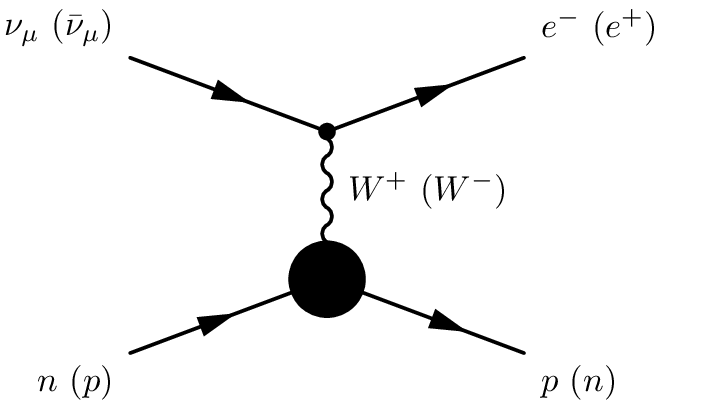}
\caption{Diagrams of non-standard neutrino-nucleon interactions for neutral-current (left) and charged-current (right) processes. }
\label{fig.feyn}
\end{figure*}
%

The neutral-current elastic (NCE) and charged-current quasi-elastic (CCQE) scattering of neutrinos with nucleons and nuclei constitute examples of fundamental electroweak interactions within the Standard Model (SM)~\cite{LlewellynSmith:1971zm}, which, despite their relative simplicity, are presently not well-understood. The first attempts of experimentally measuring the cross sections of the latter processes resulted in a  discrepancy~\cite{Cline:1976in,Entenberg:1979wc,Ahrens:1986xe} with the predictions of the widely used relativistic Fermi gas (RFG) model~\cite{Smith:1972xh,Horowitz:1993rj,Barbaro:1996vd,Alberico:1997vh,Alberico:1997rm}. There has been much effort towards quantifying this disagreement (between theory and experiment), mainly in terms of the nucleon electromagnetic form factors at intermediate energies~\cite{Ernst:1960zza,Alberico:2008sz}, while other works focus on the study of the potential contribution of the strange components of the hadronic current~\cite{Garvey:1992qp,Garvey:1992cg,Garvey:1993sg,Acha:2006my}. 

Over a decade ago, an anomalous excess of events has been reported by the LSND experiment in searching for $\nu_\mu \to \nu_e$ oscillations~\cite{Athanassopoulos:1996jb}.
Recently, in the MiniBooNE neutrino-nucleon scattering experiment, an unexplained excess of electronlike events, $\Delta N=128.8 \pm 20.4 \pm 38.3$, has been observed in the reconstructed neutrino energy range $ 200 \leq E_\nu \leq 475$~MeV~\cite{AguilarArevalo:2008rc,AguilarArevalo:2010zc,AguilarArevalo:2010cx,AguilarArevalo:2013hm,Aguilar-Arevalo:2013nkf}. To interpret these data, some authors~\cite{Bhattacharya:2011ah} argued that the RFG model is insufficient to accurately describe the neutrino-nucleon interaction in nuclei embedded in dense media~\cite{Bhattacharya:2011ah}, while other authors paid special attention to the final state interaction (FSI) effects~\cite{Nieves:2004wx,Nieves:2005rq,Golan:2012wx}. Furthermore, this anomaly has triggered the intense theoretical interest and towards its explanation several models have been proposed~\cite{Nieves:2011yp,Nieves:2011pp} including also those addressing heavy Dirac or Majorana neutrino decay~\cite{Gninenko:2009ks,Gninenko:2010pr}, the existence of sterile neutrinos~\cite{Gninenko:2012rw} and others. 

Historically, nucleons and nuclear systems have been extensively employed~\cite{Kosmas:1992yxv,Umino:1994wu} as microlaboratories for exploring open neutrino properties through charged-~\cite{Leitner:2008ue} and neutral-current interaction processes~\cite{Leitner:2006sp}. The latter involve both the vector and the axial vector components of the weak interactions~\cite{Giannaka:2015sta,Giannaka:2015zga}. Thus, such probes are helpful for investigating the fundamental interactions of neutrinos with other elementary particles at low, intermediate and high energies~\cite{Leitner:2006ww}. For the case of neutral-current coherent elastic neutrino scattering off complex nuclei, a detailed analysis focusing on possible alterations of the expected event rates due to the existence of non-standard interactions (NSI) was performed in our previous works~\cite{Papoulias:2013gha,Papoulias:2015vxa,Papoulias:2015iga}.

Motivated by the latter, in this paper, we explore the possibility of probing exotic neutrino processes in the relevant experiments. Thus, within the framework of NSI~\cite{Barger:1991ae} (for a review, see Refs.~\cite{Ohlsson:2012kf,Miranda:2015dra}) we consider the NCE scattering of the following neutrino-nucleon reactions
\begin{eqnarray}
\label{nu-nucleon-reac-1} 
\nu_\alpha ({\bar \nu}_\alpha) + n \rightarrow & \nu_\beta ({\bar \nu}_\beta) + n 
\, ,
\\
\nu_\alpha ({\bar \nu}_\alpha) +  p  \rightarrow &\nu_\beta  ({\bar \nu}_\beta) + p
\, ,
\label{nu-nucleon-reac-2}
\end{eqnarray}
where $\alpha, \beta = \{e,\mu,\tau\}$ denote the neutrino flavour. Specifically, for the case of the MiniBooNE processes, we concentrate on the channels where $\alpha = \mu$ and $\beta = e$ (for NSI scattering involving tau neutrinos, see Refs.~\cite{Rashed:2012bd,Rashed:2013dba}). In the present study, the magnitude of the proposed novel interactions is given in terms of the adopted NSI nucleon form factors as functions of the four-momentum transfer. In our effort to explore potential NSI neutrino-nucleon interactions, as a first step, we focus only on the NSI Feynman diagrams depicted in Fig.~\ref{fig.feyn} where, for completeness, the corresponding diagrams involving charged-current (CC) processes are also included.

\section{Neutral-Current Non-standard Neutrino-Nucleon Interactions}

In the present work, the assumed NSI operators are effective four-fermion operators of the form~\cite{Ohlsson:2012kf}
\begin{equation}
\mathcal{O} = \left(\bar{f}_1 \gamma^\mu P f_2\right) \left(\bar{f}_3 \gamma_\mu P f_4\right) + \mathrm{h.c.} \, ,
\label{four-fermion NSI operator}
\end{equation}
with $f_i$, $i = 1,2,3,4$  being the SM fermion fields and $P = \{L,R\}$
denoting left- and right-handed projectors. Specifically, for neutral-currents one has neutrino-induced NSI with matter of the form~\cite{Barger:1991ae}
\begin{equation}
\mathcal{O}_{\alpha \beta}^{f P} = \left(\bar{\nu}_\alpha \gamma^\mu L \nu_\beta \right) \left(\bar{f} \gamma_\mu P f\right) + \mathrm{h.c.} \, ,
\end{equation}
with $f$ denoting a first generation quark $q=\{u,d\}$. 

\subsection{Neutrino-nucleon cross sections within and beyond the SM}

The calculations of the neutrino-nucleon cross sections start by writing down the nucleon matrix elements of the processes~(\ref{nu-nucleon-reac-1}),(\ref{nu-nucleon-reac-2}) in the usual $V-A$ form, as
\begin{equation}
\begin{aligned}
\mathcal{M}=  & \frac{iG_F}{2 \sqrt{2}}  j_{\mu} \langle \mathcal{N} \vert J^{\mu}_Z \vert \mathcal{N} \rangle \\
= &\frac{iG_F}{2 \sqrt{2}}  \bar{\nu}_{\alpha}\gamma_\mu \left( 1 - \gamma_5 \right) \nu_{\beta}
\langle \mathcal{N} \vert J_{Z}^{\mu} \vert \mathcal{N} \rangle\, ,
\end{aligned}
\end{equation}
where $j_{\mu}$ denotes the leptonic neutral-current, $G_F$ the Fermi coupling constant and $\vert \mathcal{N}\rangle$ represents the nucleon wavefunction. In the latter expression, $\langle \mathcal{N} \vert J_{Z}^{\mu} \vert \mathcal{N} \rangle $ is the hadronic matrix element that is (after neglecting the second class currents and the contribution of the pseudoscalar component) expressed in terms of the known nucleon form factors as~\cite{Kosmas:1996fh}
\begin{equation}
\begin{aligned}
\langle \mathcal{N} \vert J_{Z}^{\mu} \vert \mathcal{N} \rangle = \langle \mathcal{N} \vert F_1^{\mathrm{NC}:p(n)} (Q^2) + & F_2^{\mathrm{NC}:p(n)} (Q^2) \frac{i \sigma^{\mu \nu}q_{\nu}}{2 m_{\mathcal{N}}} \\
+& F_A^{\mathrm{NC}:p(n)}(Q^2) \gamma^{\mu} \gamma^5 \vert \mathcal{N} \rangle\, .
\end{aligned}
\label{eq.ME}
\end{equation}
In Eq.~\ref{eq.ME}, $F_1^{\mathrm{NC}:p(n)} (Q^2)$, $F_2^{\mathrm{NC}:p(n)} (Q^2)$, $F_A^{\mathrm{NC}:p(n)} (Q^2)$ stand for the Dirac, Pauli and axial vector weak neutral-current form factors, respectively, for protons ($p$) or neutrons ($n$)~\cite{Alberico:2001sd}.

Relying on the above nucleon matrix elements, within the relativistic Fermi gas (RFG) model, the SM differential cross section of reactions~(\ref{nu-nucleon-reac-1}) and (\ref{nu-nucleon-reac-2}) for incoming (anti)neutrino energy $E_\nu$ has been written as~\cite{LlewellynSmith:1971zm}
\begin{equation}
\frac{d \sigma}{dQ^2}= \frac{G_F^2 Q^2}{2 \pi E_\nu^2} \left[ A(Q^2) \pm B(Q^2) W + C(Q^2) W^2 \right]\, .
\label{dsdQ}
\end{equation}
In the above expression, the plus (minus) sign accounts for neutrino (antineutrino) scattering while the momentum-depended function, $W(Q^2)$, reads~\cite{Garvey:1992cg}
\begin{equation}
W= \frac{4 E_\nu}{m_{\mathcal{N}}} - \frac{Q^2}{m_{\mathcal{N}}^2}\, ,
\end{equation}
where the four-momentum transfer is defined in terms of the nucleon recoil energy $T_{\mathcal{N}}$, as
\begin{equation}
q^2 =q_\mu q^\mu=-Q^2 = -2 m_{\mathcal{N}} T_{\mathcal{N}} \, .
\end{equation}
For the nucleon mass, $m_{\mathcal{N}}$, we assume the value $m_p \approx m_n=m_{\mathcal{N}} = 0.938$~GeV.

Before proceeding to the cross sections calculations, for the reader's convenience, we provide below some significant details on the aforementioned expressions $A$, $B$ and $C$ that depend on the form factors $F_i^{\mathrm{NC}:p(n)}$, $i=1,2,A$.
The functions $A(Q^2)$, $B(Q^2)$ and $C(Q^2)$ are defined as~\cite{Garvey:1993sg}
\begin{equation}
\begin{aligned}
A(Q^2) = \frac{1}{4} \Biggl\{  & \left(F_A^{\mathrm{NC}:p(n)} \right)^2 (1+\tau) \\
& - \left[ \left(F_1^{\mathrm{NC}:p(n)} \right)^2 - \tau  \left(F_2^{\mathrm{NC}:p(n)} \right)^2 \right] (1-\tau)\\
& + 4 \tau F_1^{\mathrm{NC}:p(n)} F_2^{\mathrm{NC}:p(n)}\Biggr\}\, ,\\
\end{aligned}
\end{equation}
\begin{equation}
B(Q^2) = - \frac{1}{4} F_A^{\mathrm{NC}:p(n)} \left(F_1^{\mathrm{NC}:p(n)} +F_2^{\mathrm{NC}:p(n)} \right)\, ,
\end{equation}
\begin{equation}
\begin{aligned}
C(Q^2) = \frac{m_{\mathcal{N}}^2}{16 Q^2} \Biggl[& \left(F_A^{\mathrm{NC}:p(n)} \right)^2 + \left(F_1^{\mathrm{NC}:p(n)} \right)^2 \\
& + \tau \left(F_2^{\mathrm{NC}:p(n)} \right)^2 \Biggr]\, ,
\end{aligned}
\end{equation}
where their explicit $Q^2$ dependence has been suppressed and $\tau =Q^2/4 m_{\mathcal{N}}^2$.

In principle, the electromagnetic Dirac and Pauli form factors are written in terms of the well-known electric (E) and magnetic (M) form factors as follows~\cite{Alberico:2001sd}
\begin{equation}
\begin{aligned}
F_1^{\mathrm{EM}:p(n)} =& \frac{G_\mathrm{E}^{p(n)}(Q^2) + \tau G_\mathrm{M}^{p(n)}(Q^2)}{1+\tau}\, ,\\
F_2^{\mathrm{EM}:p(n)} =& \frac{G_\mathrm{M}^{p(n)}(Q^2) - \tau G_\mathrm{E}^{p(n)}(Q^2)}{1+\tau}\, .\\
\end{aligned}
\end{equation}
In this work, the magnetic form factors are parametrised as~\cite{Alberico:2008sz}
\begin{equation}
\frac{G_\mathrm{M}^{p(n)}}{\mu_{p(n)}} = \frac{1 + a^\mathrm{M}_{p(n),1} \tau}{1 + b^\mathrm{M}_{p(n),1} \tau  + b^\mathrm{M}_{p(n),2} \tau^2 + b^\mathrm{M}_{p(n),3} \tau^3}\, ,
\end{equation}
where $\mu_{p(n)}$ denotes the proton (neutron) magnetic moment. The proton electric form factor in a similar manner can be cast in the form~\cite{Balasi:2011zz}
\begin{equation}
G_\mathrm{E}^{p(n)} = \frac{1 + a^\mathrm{E}_{p(n),1} \tau}{1 + b^\mathrm{E}_{p(n),1} \tau  + b^\mathrm{E}_{p(n),2} \tau^2 + b^\mathrm{E}_{p(n),3} \tau^3}\, ,
\end{equation}
(for the fit parameters, $a^\mathrm{M(E)}_{p(n),1}$ and $b^\mathrm{M(E)}_{p(n),j}$, $j=1,2,3$, see Ref.~\cite{Alberico:2008sz}). The electric neutron form factor, $G_\mathrm{E}^{n}$, is expressed through the Galster-like parametrisation, as
\begin{equation}
G_\mathrm{E}^{n}(Q^2) = \frac{\lambda_1 \tau}{1 + \lambda_2 \tau} G_D(Q^2)\, ,
\end{equation}
with $\lambda_1 = 1.68$ and $\lambda_2 = 3.63$.
%

\subsection{NSI nucleon form factors}

As it is well known, within the SM, the weak NC Dirac and Pauli form factors are written in terms of the electromagnetic current form factors $F_i^{\mathrm{EM}}$, $i=1,2$ [assuming the conserved vector current (CVC) theory]~\cite{Alberico:2001sd}. In the present work, we furthermore consider additional contributions originating from NSI that enter through the vector-type form factors $\varepsilon_{\mu e}^{q V}(Q^2)$. In our parametrisation, the latter are written in terms of the fundamental NSI neutrino-quark couplings $\epsilon_{\mu e}^{u V}$ ($\epsilon_{\mu e}^{d V}$) for $u$ ($d$) quarks discussed in Refs.~\cite{Papoulias:2013gha,Papoulias:2015vxa,Papoulias:2015iga}, and they take the form
\begin{equation}
\begin{aligned}
\varepsilon_{\mu e}^{p V}(Q^2) = (2 \epsilon_{\mu e}^{u V} + \epsilon_{\mu e}^{d V}) G_D(Q^2)\, ,\\
\varepsilon_{\mu e}^{n V}(Q^2) = (\epsilon_{\mu e}^{u V} + 2\epsilon_{\mu e}^{d V}) G_D(Q^2)\, .\\
\end{aligned}
\label{NSI.ff}
\end{equation}
In the spirit of previous studies which consider the strangeness of the nucleon~\cite{Garvey:1992cg,Garvey:1993sg}, the above NSI form factors may have the same momentum dependence as those of the SM ones. Thus, the function $G_D(Q^2)$ is assumed to be of dipole type
\begin{equation}
G_D = \left( 1 + \frac{Q^2}{M_V^2} \right)^{-2}\, ,
\end{equation}
(for the vector mass a commonly used value is $M_V = 0.843$~GeV). A dipole approximation for $G_D(Q^2)$, apart from providing the appropriate momentum dependence, ensures also that the event rate coming out of NSI has the correct behaviour at high energies \cite{Garvey:1992qp} 

%
\begin{figure}[t]
\centering
\includegraphics[width=0.5 \textwidth]{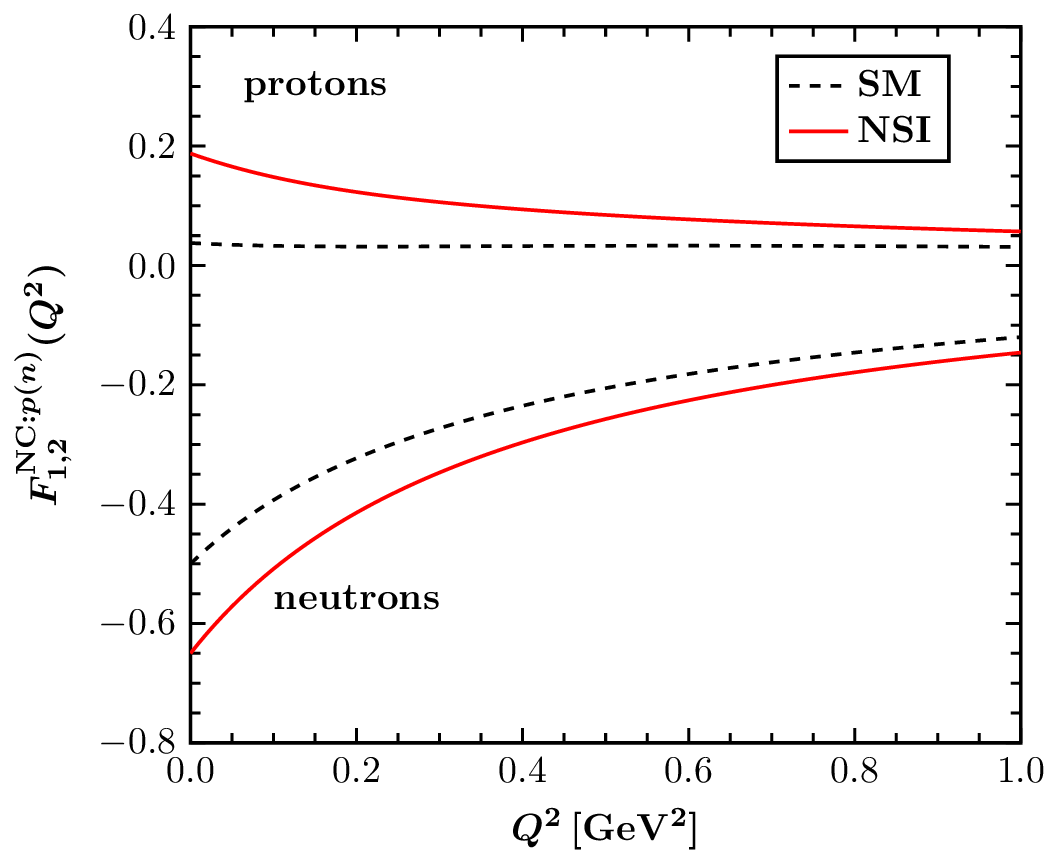}
\caption{Comparison of the SM and NSI nucleon form factors employed in the present study (for details, see the text). }
\label{fig.NSI-FF}
\end{figure}
%

Then, the weak neutral-current nucleon form factors for protons (plus sign) and neutrons (minus sign) employed in our present calculations read
\begin{equation}
\begin{aligned}
F_{1,2}^{\mathrm{NC}:p(n)}(Q^2) = &  \frac{\tau_3}{2} \left[F_{1,2}^{\mathrm{EM}:p}(Q^2) 
- F_{1,2}^{\mathrm{EM}:n}(Q^2) \right] \\
& - 2 \sin^2 \theta_W F_{1,2}^{\mathrm{EM}:p(n)} (Q^2) \\
& -\frac{1}{2} F_{1,2}^{\mathrm{s}:p(n)}(Q^2)\\
& + \tau_3 \varepsilon_{\mu e}^{p(n) V}(Q^2) \, .
\end{aligned}
\end{equation} 
In the latter expression the isoscalar form factors $F_{1,2}^{\mathrm{s}:p(n)}$ account for potential contributions to the electric charge and the magnetic moment
of the nucleon due to the presence of strange quarks (in our convention the isospin index $\tau_3$ is +1 for proton and -1 for neutron scattering). However, throughout our calculations, on the basis of the recent results from the HAPPEX experiment~\cite{Acha:2006my}, we take $F_{1,2}^{\mathrm{s}:p(n)} =0$ (see also Ref.~\cite{Nieves:2005rq}).
Note that, for low momentum transfer, the form factors discussed in Refs.~\cite{Papoulias:2013gha,Papoulias:2015vxa} are recovered. The effect of NSI on the form factors is illustrated graphically in Fig.~\ref{fig.NSI-FF}, where typical values have been adopted for the NSI parameters, i.e.  $\epsilon_{\mu e}^{uV}=\epsilon_{\mu e}^{dV}= 0.05$ (apparently strange quark contributions have no impact in this case). Within the present formalism, the new nucleon form factors are rather sensitive to NSI, even for small values of the fundamental model parameters, especially for low momentum transfer.
 
For the case of the axial form factor $F_A^{\mathrm{NC}:p(n)}$, we correspondingly employed~\cite{Leitner:2008ue}
\begin{equation}
F_A^{\mathrm{NC}:p(n)}(Q^2) = \frac{1}{2} (\tau_3 g_A + g_A^{\mathrm{s}}) \left( 1 + \frac{Q^2}{M_A^2} \right)^{-2}\, .
\label{NSI-FA}
\end{equation}
Here, we used the static axial vector coupling, $g_A=-1.267$  (it is usually determined through neutron beta decay). For the strange quark contribution to the nucleon spin, we adopt the static value $2F_A^{\mathrm{s}}(0) =g_A^{\mathrm{s}} \pm 0.07$ with $g_A^{\mathrm{s}}=-0.15$, while for the axial mass we take $M_A=1.049$~GeV (i.e. fit II of Ref.~\cite{Garvey:1992cg}). As has been recently discussed in Ref.~\cite{Nieves:2011yp} this set of values is fully compatible with the MiniBooNE data, even though a large value of $M_A = 1.35$~GeV was reported in Ref.~\cite{AguilarArevalo:2010cx}. Furthermore, for simplicity, potential axial NSI form factors are neglected.

\section{Results and discussion}

%
\begin{figure}[t]
\centering
\includegraphics[width=0.49 \textwidth]{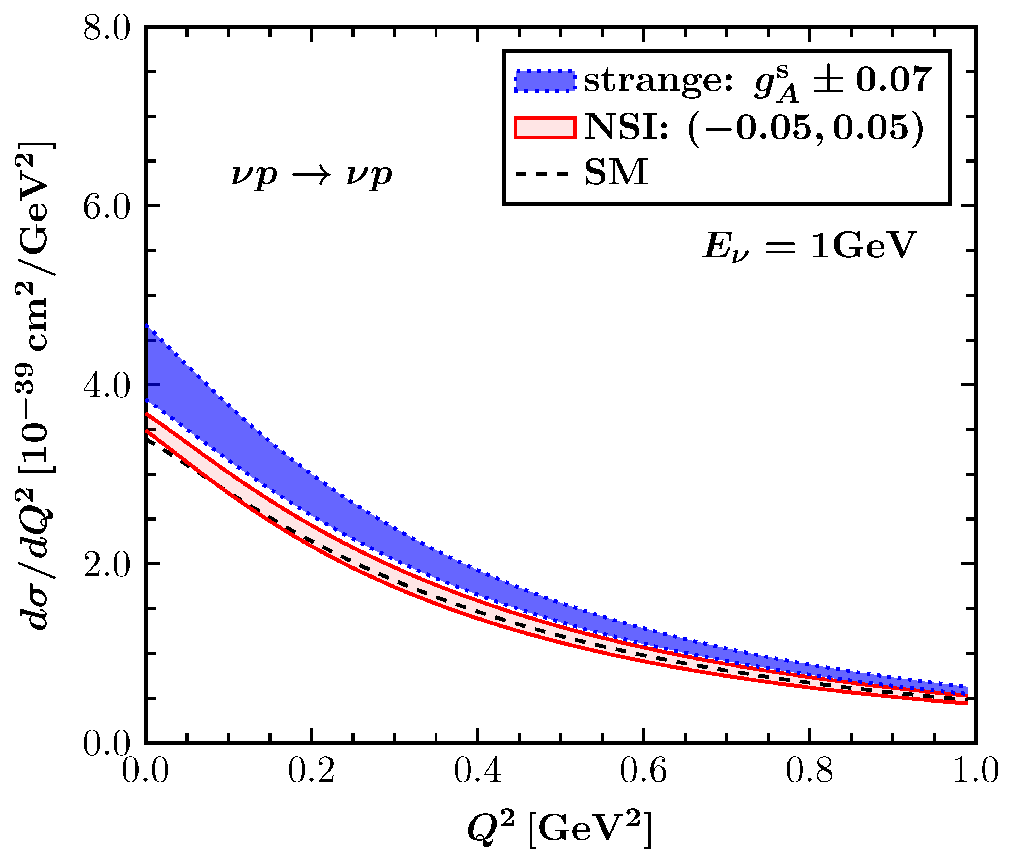}
\includegraphics[width=0.49 \textwidth]{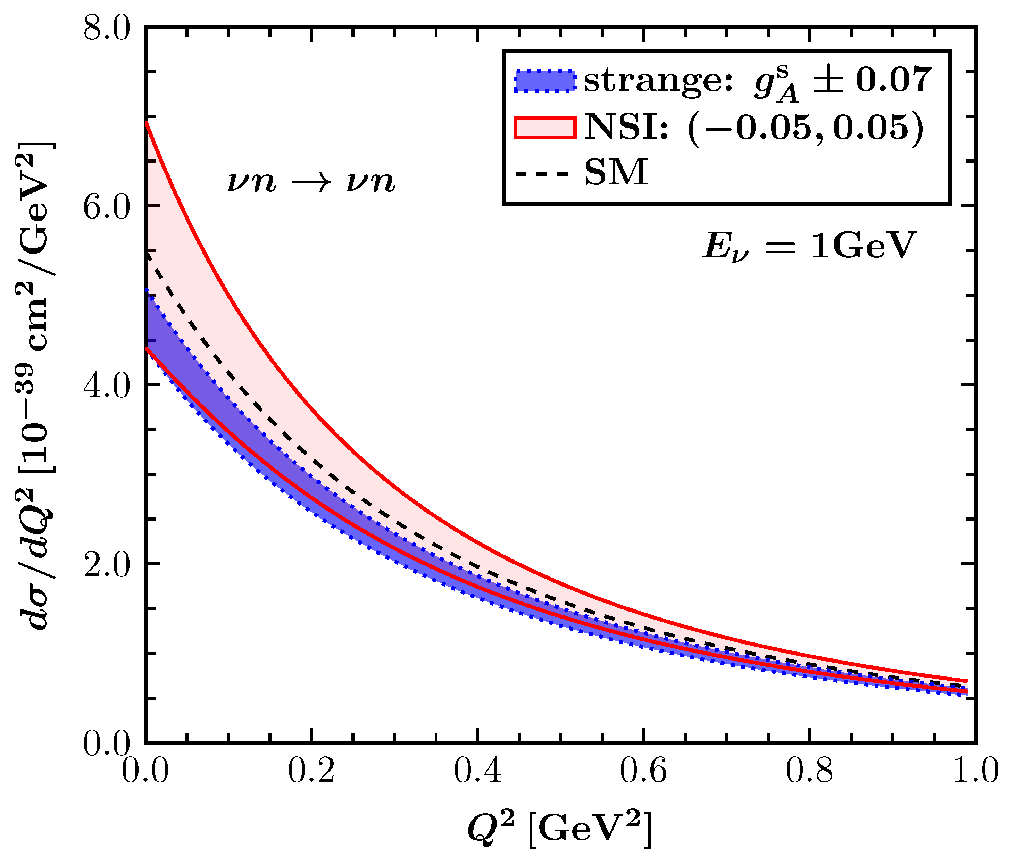}
\caption{Differential cross section with respect to the momentum transfer for SM, strange quark and NSI, for $\nu p \to \nu p$ (upper panel) and $\nu n \to \nu n$ scattering (lower panel). For details, see the text.}
\label{fig.dsdQ}
\end{figure}
%
%
\begin{figure}[t]
\centering
\includegraphics[width=0.49 \textwidth]{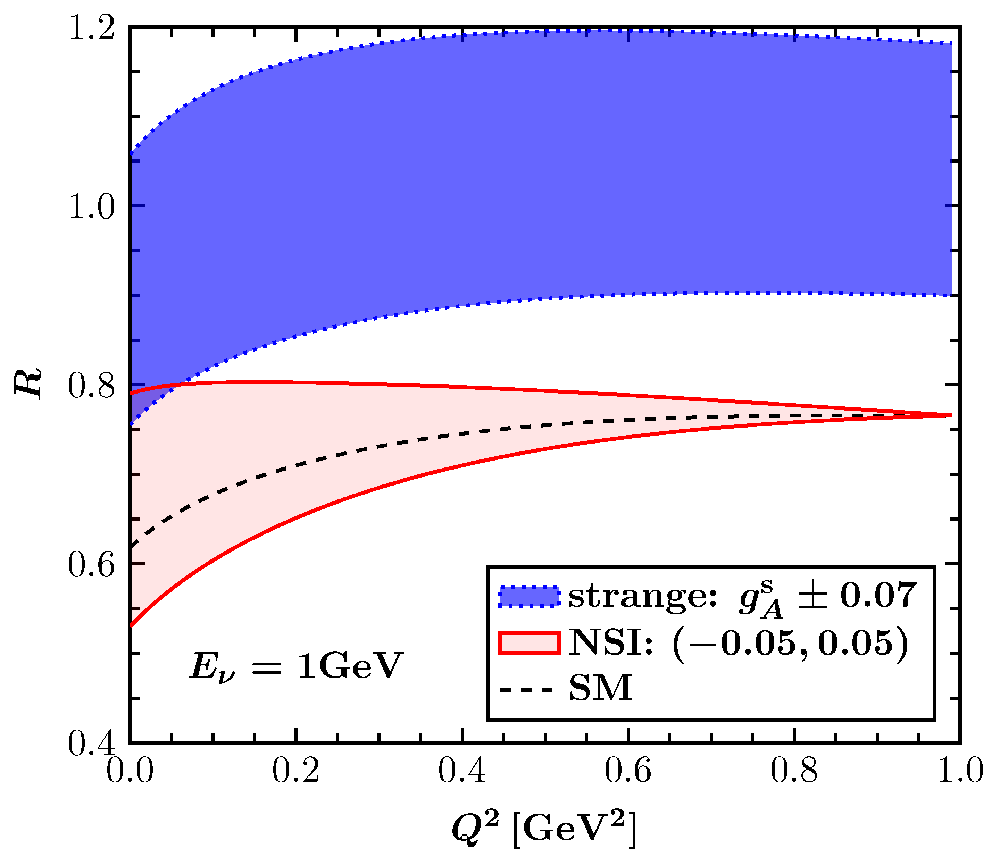}
\caption{Momentum variation of the cross sections ratio for SM, strange quark and NSI neutrino-nucleon scattering.}
\label{fig.ratio}
\end{figure}
%

At first, we calculate the differential cross sections of Eqs.~(\ref{dsdQ}) for the SM ($\alpha = \beta$) weak NCE scattering processes~(\ref{nu-nucleon-reac-1}) and (\ref{nu-nucleon-reac-2}) as well as for the NSI ones ($\alpha \neq \beta$), based on Eqs.~(\ref{NSI.ff})-(\ref{NSI-FA}) (neglecting potential strange quark contributions, i.e. $F_{1,2}^{\mathrm{s}:p(n)}=0$ and $g_A^{\mathrm{s}}=0$). The corresponding results are demonstrated in Fig.~\ref{fig.dsdQ} for $\nu p \to \nu p$ scattering (upper panel) and $\nu n \to \nu n$ scattering (lower panel). For the sake of comparison, the bands of axial vector strange quark contributions calculated within $g_A^\mathrm{s} \pm 0.07$ and those of NSI contributions within the range $\epsilon_{\mu e}^{q V} = (-0.05,0.05)$ with $q=\{u,d\}$, are also depicted. One sees that the resulting strange quark contributions indicate almost equal cross sections for proton and neutron scattering, while the presence of NSI leads to an enhancement of the cross sections for both $\nu p \rightarrow \nu p$ and $\nu n \rightarrow \nu n$ scattering channels, which becomes more important at lower energies.   

From the perspective of experimental physics, it is crucial to reduce most of the background as well as beam related and systematic uncertainties. Therefore, a rather advantageous way towards determining the strange or NSI parameters is to perform measurements of the ratio of the NCE cross sections 
\begin{equation}
R=\frac{d\sigma_p /dQ^2}{d\sigma_n /dQ^2} \, .
\end{equation}
In this context, Fig.~\ref{fig.ratio} illustrates a comparison of the obtained bands for the ratio $R$ assuming neutrino-nucleon scattering in the presence of strange quarks or potential NSI. One notices that $R$ varies between 0.75 and 1.20 when axial vector strange quark contributions are taken into account, while for the case of NSI the ratio is significantly lower, i.e. it lies between 0.55 and 0.8. It is furthermore shown that, unlike the strange quark case, the SM result for $R$ lies within the predicted NSI band.

\begin{figure}[t]
\centering
\includegraphics[width=0.5\textwidth]{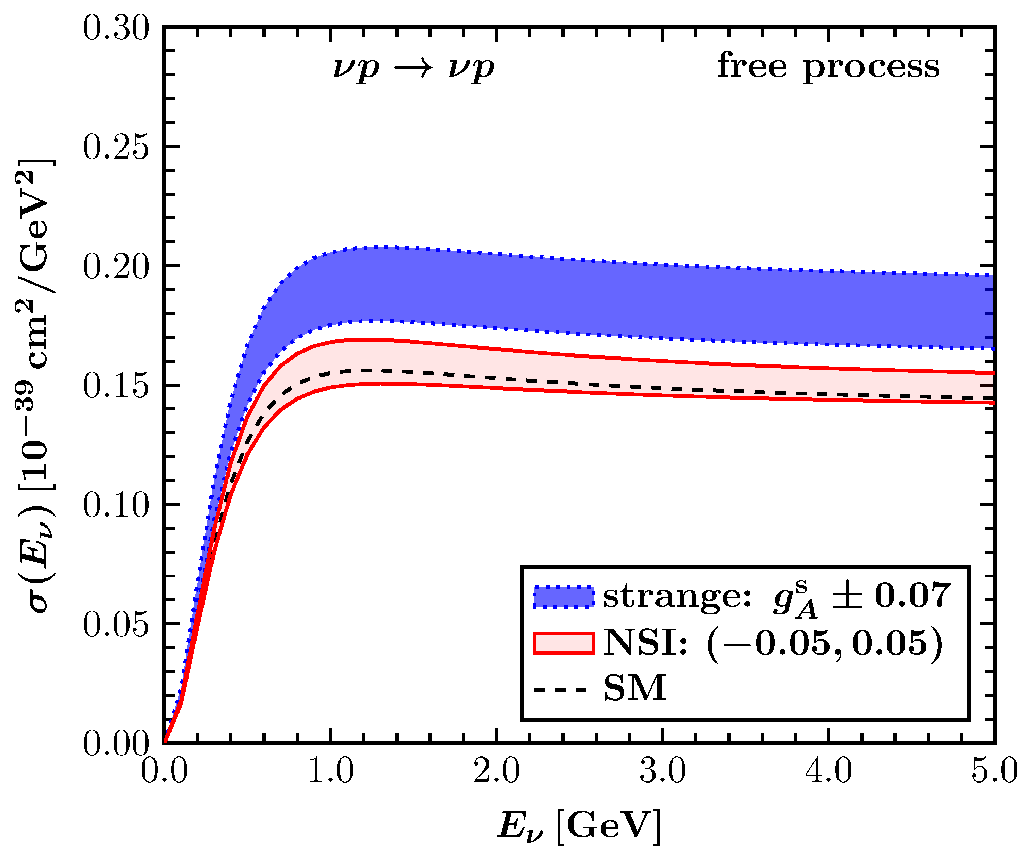}
\includegraphics[width=0.5\textwidth]{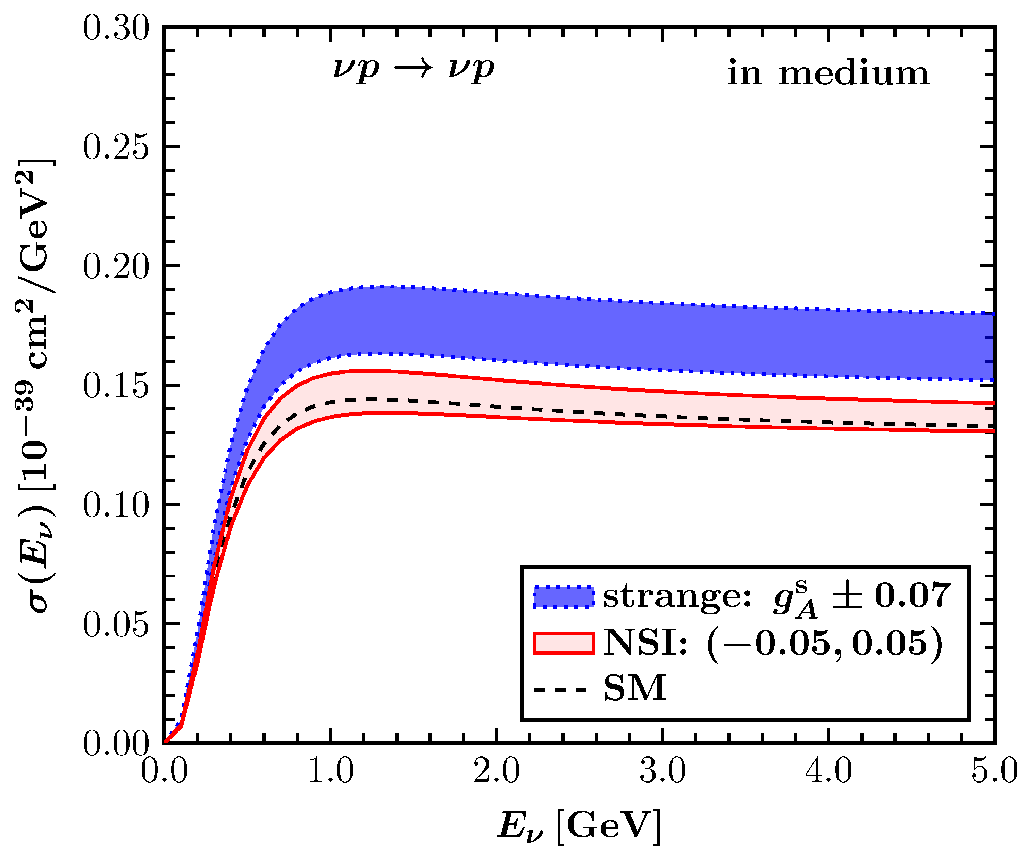}
\caption{Total integrated $\nu p \to \nu p$ scattering cross section as a function of the incoming neutrino energy due to SM, strange quark and NSI scattering. The results refer to scattering on free protons (upper panel) and scattering on bound protons i.e. assuming nuclear effects at a $^{12}$C detector (lower panel).}
\label{fig.tot-crossec-protons}
\end{figure}

\begin{figure}[t]
\centering
\includegraphics[width=0.5\textwidth]{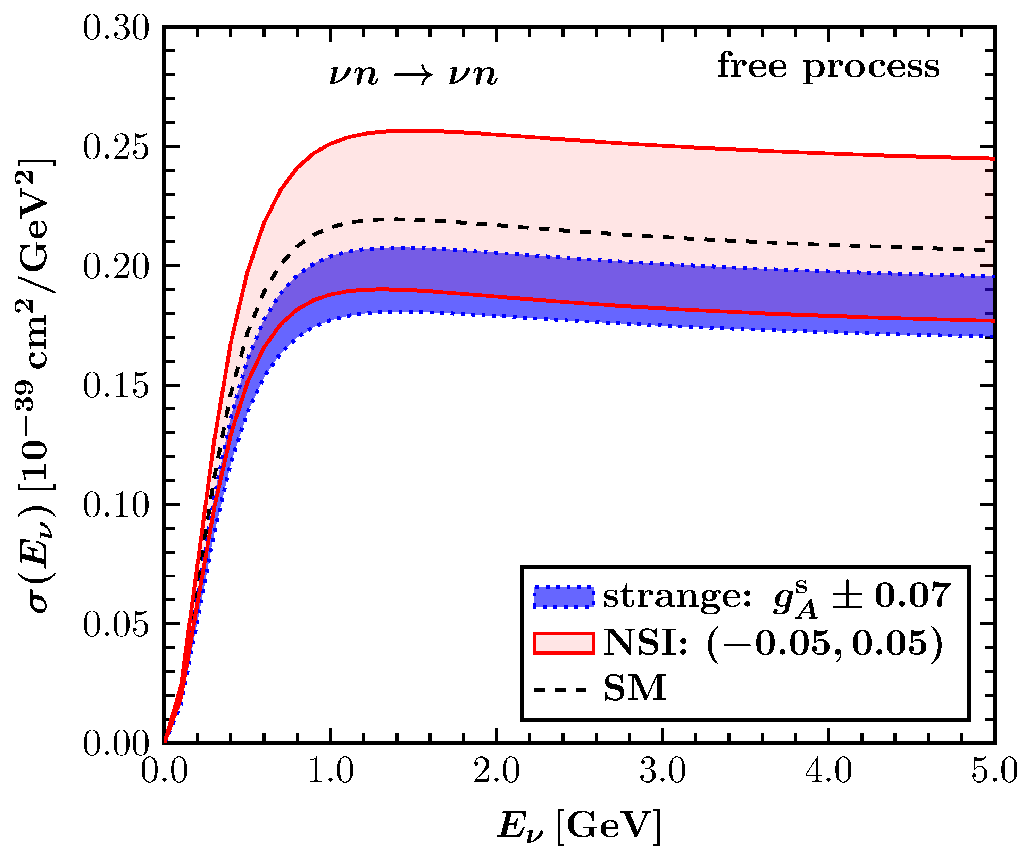}
\includegraphics[width=0.5\textwidth]{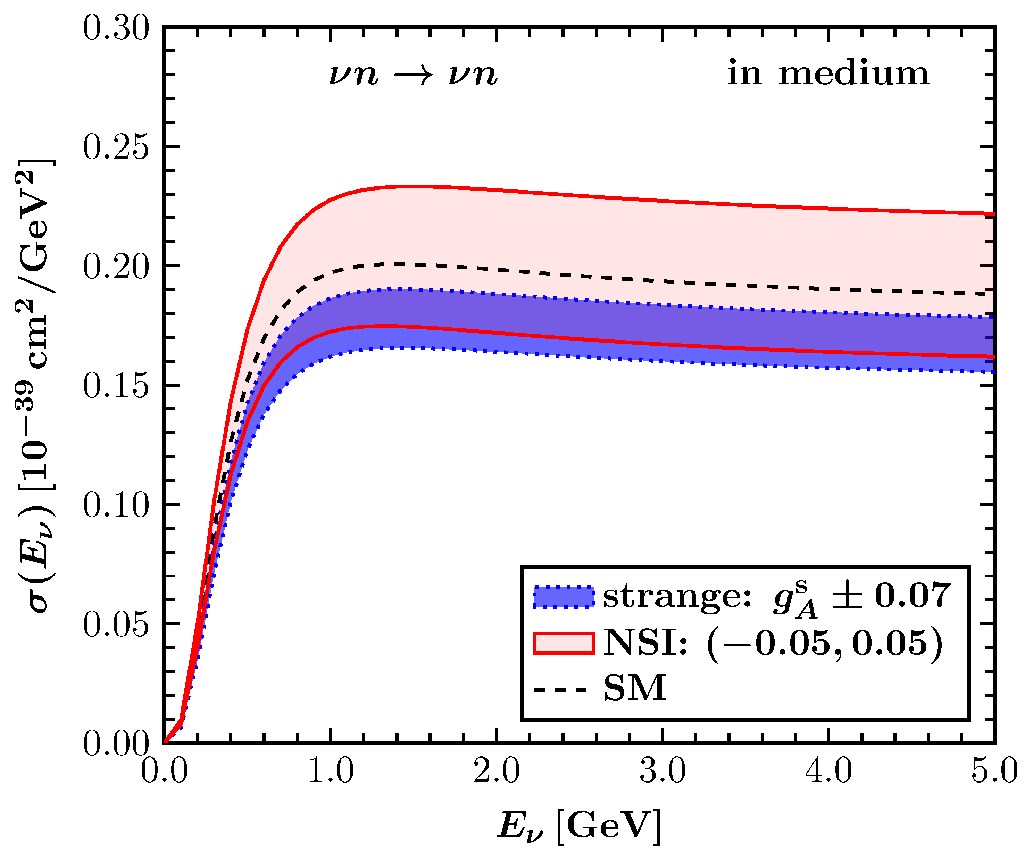}
\caption{Same as Fig.~\ref{fig.tot-crossec-protons} but for $\nu n \to \nu n$ scattering.}
\label{fig.tot-crossec-neutrons}
\end{figure}

Focusing on the relevant experiments, in order to perform reliable calculations, important effects that originate from the Pauli principle must be taken into account (Fermi motion of the initial nucleons). Specifically, for the case of nucleons bound in nuclear matter, within the Fermi gas model, it is adequate to multiply the free cross section given in Eq.~(\ref{dsdQ}) with a suppression factor $S(Q^2)$. The latter  accounts for the Pauli blocking effect on the final nucleons in a local density approximation~\cite{Leitner:2006sp}, and is given by the expression~\cite{LlewellynSmith:1971zm}
\begin{equation}
S(Q^2) = 1-D(Q^2)/N \, ,
\end{equation}
where $N$ is the number of neutrons of the nuclear target. Assuming a carbon target, $^{12}$C, $D(Q^2)$ takes the form
\begin{equation}
D(Q^2) = \begin{cases}
A/2 \left[1-3/4 \frac{ \vert\boldsymbol{q} \vert }{p_F} + 1/24 \left(\frac{\vert \boldsymbol{q} \vert }{p_F} \right) ^3 \right], & \vert \boldsymbol{q}\vert <2 p_F\, , \\
0, &  \vert \boldsymbol{q}\vert>2 p_F\, , 
\end{cases}
\end{equation}
with the Fermi momentum $p_F = 0.220$~GeV (for its definition, see Refs.~\cite{Leitner:2006ww,Leitner:2008ue}) and $\vert \mathbf{q} \vert$ being the magnitude of the three-momentum transfer. Within this framework, the total cross section may be evaluated through numerical integration of the differential cross section (\ref{dsdQ}) as
\begin{equation}
\sigma(E_\nu) = \int_0^{Q_{\mathrm{max}}^2(E_\nu)} S(Q^2) \frac{d\sigma}{dQ^2}  \, dQ^2 \, .
\end{equation}
For NCE scattering, the kinematics of the process provide the approximate upper limit of the momentum transfer $Q^2$, as 
\begin{equation} 
Q_{\mathrm{max}}^2(E_\nu) = \frac{4 m_{\mathcal{N}} E_\nu^2}{m_{\mathcal{N}} + 2E_\nu} \, .
\end{equation}
The obtained results are demonstrated in Fig.~\ref{fig.tot-crossec-protons} for protons and in Fig.~\ref{fig.tot-crossec-neutrons} for neutrons for the case of (i) free nucleon
scattering, that is, on a hydrogen atom (upper panel), and (ii) medium
scattering, that is, for bound nucleons within a carbon
atom (lower).
As expected, for both SM and NSI, the obtained integrated cross section for the $\nu n$ process is much larger than that of the $\nu p$ reaction. Apparently, one also notices that the NCE cross sections, for both $\nu p \to \nu p$, $\nu n \to \nu n$ scattering channels, are enhanced when potential nonzero NSI are assumed. For the case of neutrino-proton scattering, the considered axial vector strange quark effects dominate the total cross section. On the other hand, focusing on neutrino-neutron scattering, the assumed strangeness of the nucleon leads to a suppression of the total cross section. Eventually, we find that the resulting strange quark and NSI bands overlap only for $\nu n$ processes.

%
%
%
%
%
\begin{figure}[t]
\centering
\includegraphics[width=0.5\textwidth]{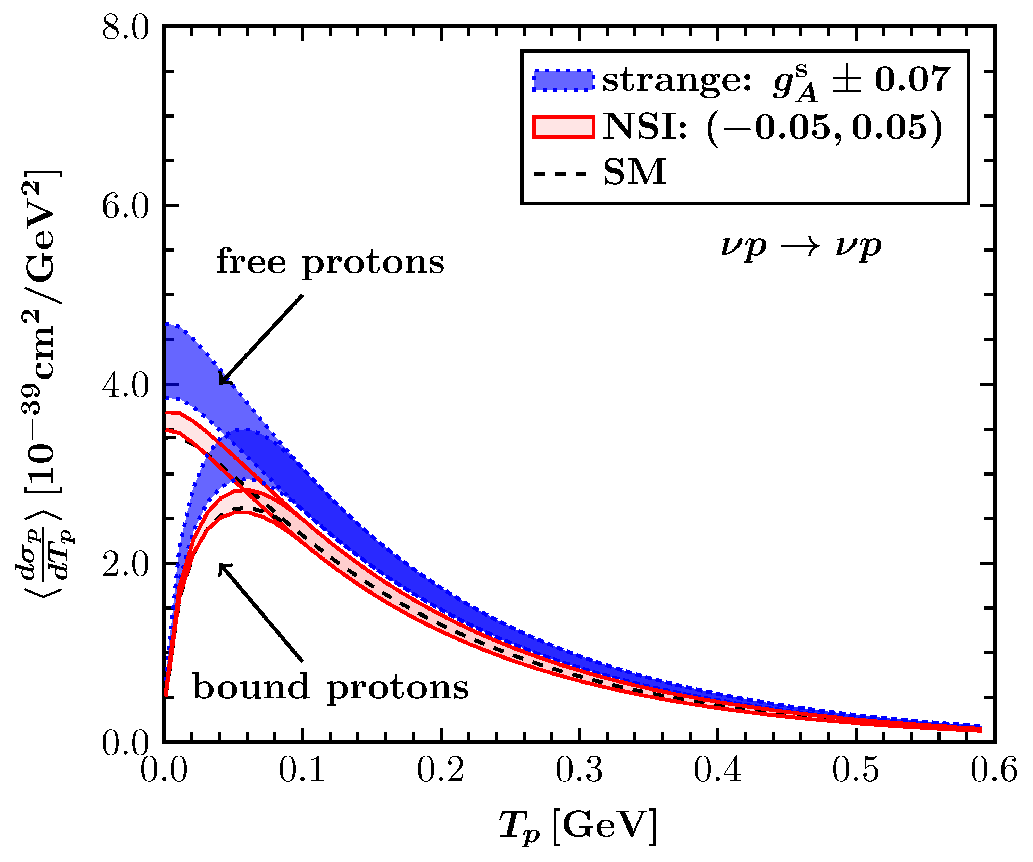}
\includegraphics[width=0.5\textwidth]{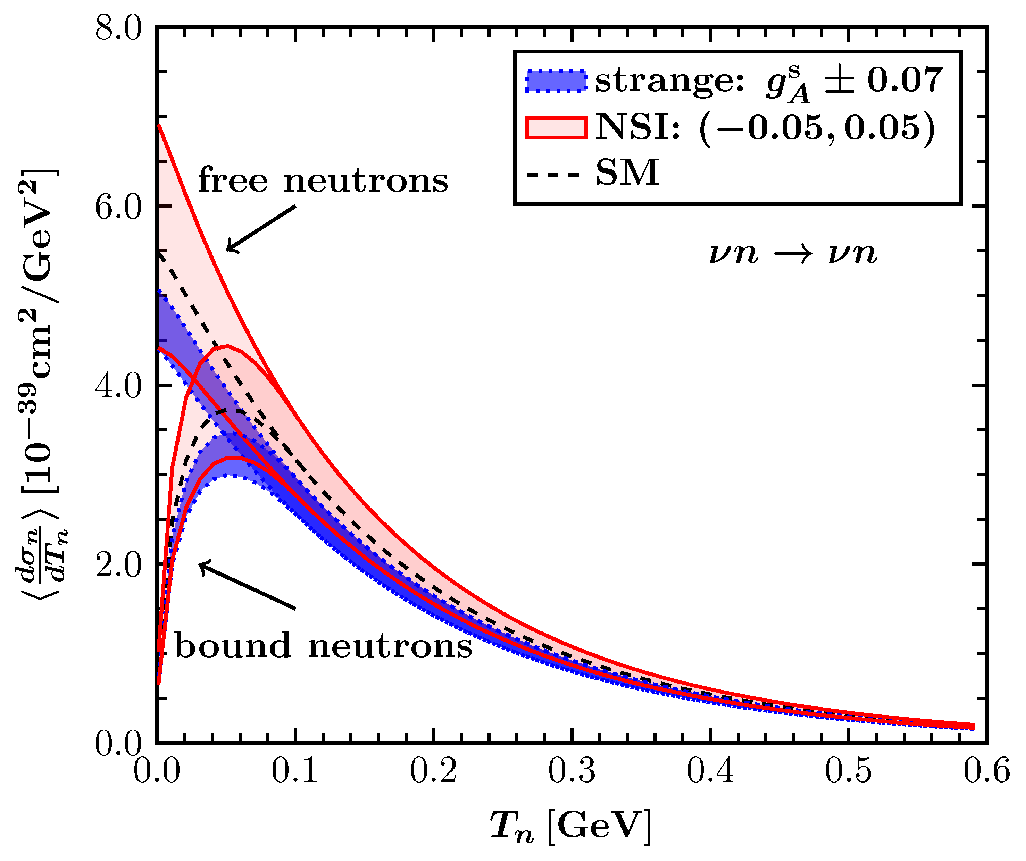}
\caption{Differential cross section as a function of the nucleon (proton or neutron) recoil energy due to SM, strange quark and NSI. Modifications due to the employed nuclear effects are illustrated and compared with the case of scattering on free nucleons.}
\label{fig.conv-NSI-crossec}
\end{figure}

%

At this point, we find it interesting to focus our discussion on the MiniBooNE experiment and apply the addressed NSI model. Thus, by convoluting the energy distribution of the MiniBooNE neutrino beam with the NSI cross section of Eq.~\ref{dsdQ}, we evaluate the flux-integrated differential NSI neutrino-nucleon cross sections $\langle d\sigma_{p,n}/dT_{p,n} \rangle$, through the expression
\begin{equation}
\begin{aligned}
\langle \frac{d\sigma_{p,n}}{dT_{p,n}} \rangle = \int & S(Q^2) \frac{d\sigma}{dQ^2} (E_\nu,Q^2) \Phi_\nu(E_\nu)  \\
& \times \delta\left( \frac{Q^2}{ 2 m_{\mathcal{N}}} - T_{p,n}\right)\, d E_\nu dQ^2 \, ,
\end{aligned}
\end{equation}
where the utilised muon neutrino flux, $\Phi_\nu(E_\nu)$, is normalised to unity. The results are illustrated in Fig.~\ref{fig.conv-NSI-crossec}, where it is clearly shown that the nuclear effects become important at low recoil energies. More specifically, for NCE scattering on free nucleons (e.g. for a hydrogen target) the cross section is significantly larger and constantly increasing for low nucleon recoil energies. On the contrary, for the case of bound nucleons within the carbon target material, $^{12}$C, the behaviour of the cross section changes drastically at low recoil energies and its value minimises for energies $\lesssim 100$~MeV. As for the total integrated cross sections discussed previously, our results show an overlap between the obtained strange quark and NSI bands only for the processes involving $\nu n \to \nu n$ scattering.
%
\begin{figure}[t]
\centering
\includegraphics[width=0.5 \textwidth]{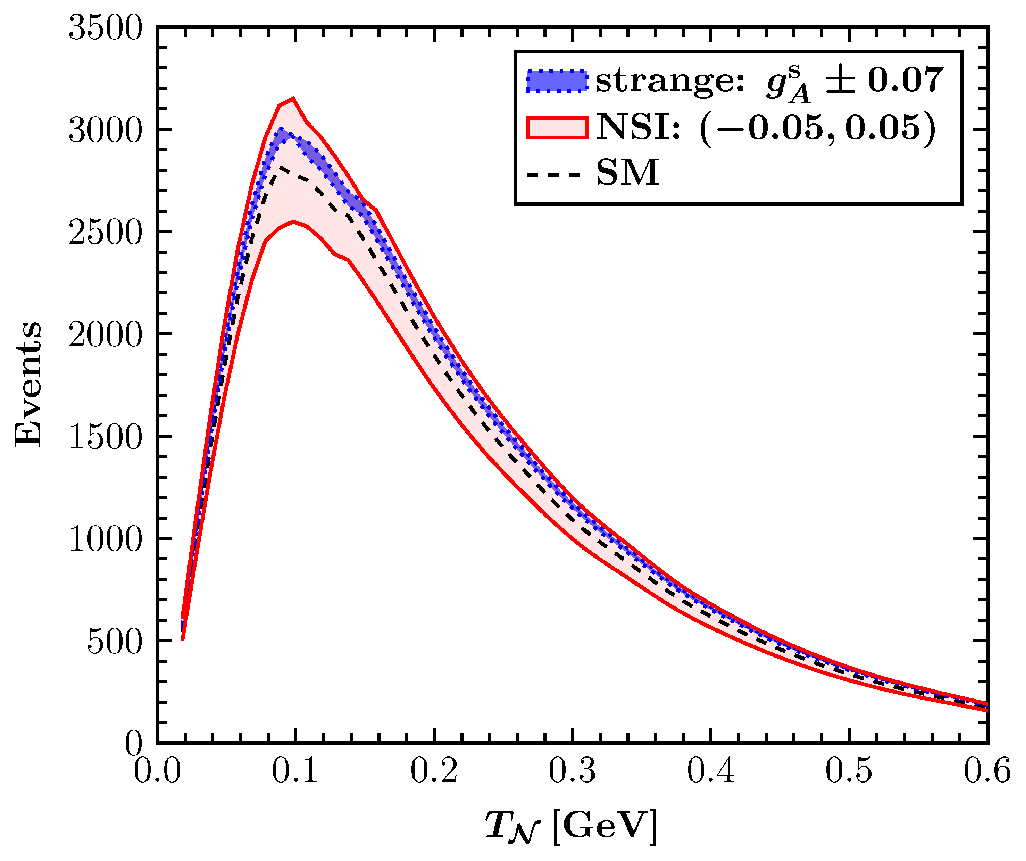}
\caption{Expected number of events as a function of the nucleon recoil energy $T_{\mathcal{N}}$ assuming contributions due to the SM, strange quark and NSI.}
\label{fig.events}
\end{figure}
%
%
\begin{figure}[t]
\centering
\includegraphics[width=0.5 \textwidth]{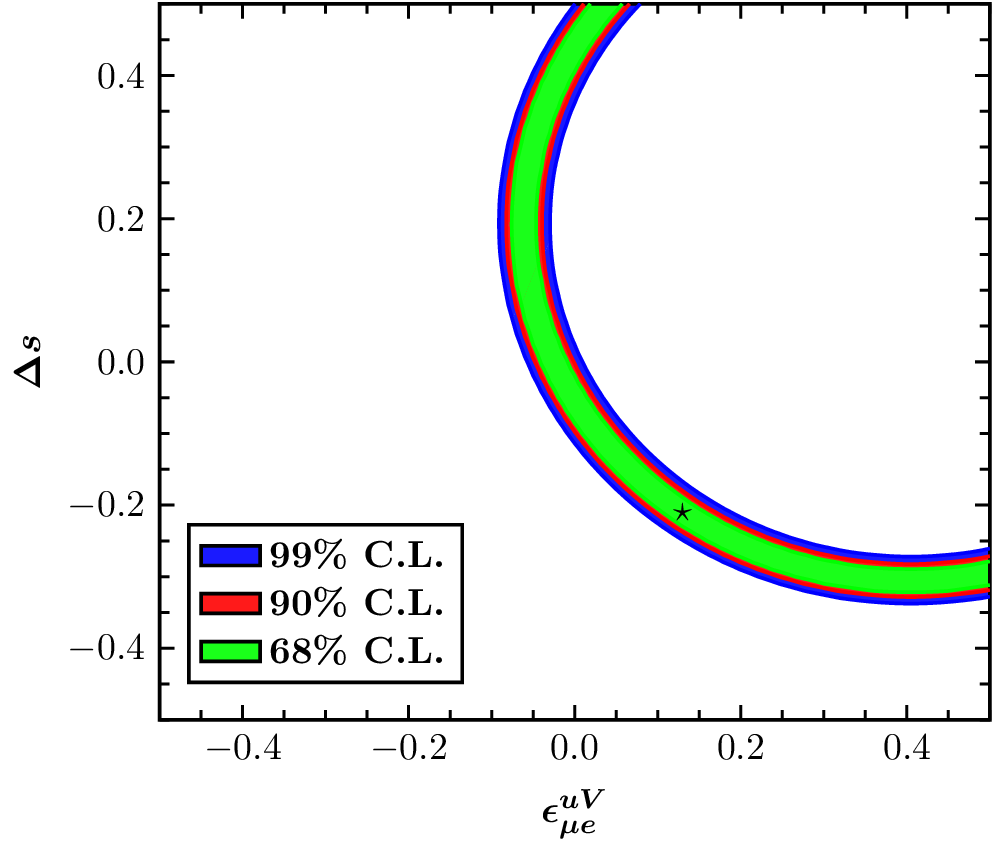}
\caption{Allowed region in the $(\epsilon_{\mu e}^{uV}-\Delta s)$ plane at MiniBooNE. The best fit point is shown by $\star$.}
\label{fig.overlap}
\end{figure}
%

We finally test the compatibility of the employed NSI scenario with recent results from the LSND and MiniBooNE experiments. To this aim, concentrating on neutrino-nucleon scattering on a mineral oil ($\mathrm{CH}_2$), the folded differential cross section reads~\cite{AguilarArevalo:2010cx}
\begin{equation}
\langle \frac{d\sigma_{\mathcal{N}}}{dT_{\mathcal{N}}} \rangle = 1/7 \langle \frac{d\sigma_p^H}{dT_p} \rangle + 3/7 \langle \frac{d\sigma_p^C}{dT_p} \rangle + 3/7 \langle \frac{d\sigma_n^C}{dT_n} \rangle \, .
\end{equation}
Then, we evaluate the number of conventional and NSI events as a function of the nucleon recoil energy, $T_{\mathcal{N}}$, by assuming a $\mathrm{CH}_2$ detector (i.e. the detector material of the LSND and MiniBooNE experiments) through the expression
\begin{equation} 
N_{\mathrm{events}}= a(T_{\mathcal{N}}) \int \mathcal{N}_N N_\mathrm{POT} \langle \frac{d\sigma_{\mathcal{N}}}{dT_{\mathcal{N}}} \rangle  dT_{\mathcal{N}} \, .
\end{equation}
In order to confront our present results with the recent MiniBooNE data, we assumed the following experimental quantities: the utilised muon neutrino flux, $\Phi_\nu(E_\nu)$, is normalised to the protons on target (POT) with $N_\mathrm{POT}$ denoting the number of POT in the data~\cite{AguilarArevalo:2010zc} and $a(T_{\mathcal{N}})$ denoting the detector efficiency taken from Ref.~\cite{Aguilar-Arevalo:2013nkf}. The number of nucleon targets in the detector is evaluated as $\mathcal{N}_N = N_A \frac{4}{3} \pi R^3 \rho_{\mathrm{oil}}$, where the density of mineral oil is $\rho_{\mathrm{oil}} = 0.845\, \mathrm{gr/cm^3}$ at $20~^oC$~\cite{AguilarArevalo:2013hm} and $N_A$ the Avogadro's number. The fiducial volume cut of the detector is adopted from Ref.~\cite{AguilarArevalo:2010cx}. Figure~\ref{fig.events} illustrates the number of events as a function of the nucleon kinetic energy, $T_{\mathcal{N}}$, obtained  within the context of the  SM,  as well as by assuming potential contributions to the rate arising from strange quarks or NSI (for a comparison with the MiniBooNE experimental results, see Ref.~\cite{AguilarArevalo:2010cx}). The obtained excess of events becomes significant for low recoil energies, reflecting the dipole character of the form factor $G_D(Q^2)$ that enters the definition of the NSI nucleon form factors given in Eq.~(\ref{NSI.ff}).

Motivated by our previous studies, we consider it interesting to estimate the sensitivity of MiniBooNE to non-standard interactions, through a $\chi^2$ fit analysis. By varying one NSI coupling at a time and by neglecting potential strange quark contributions, the minimisation of the $\chi^2(\epsilon_{\mu}^{uV})$ function provides constraints on the NSI parameters of the order of $\epsilon_{\mu e}^{dV} \approx \epsilon_{\mu e}^{dV}= 0.05$. Furthermore, in order to explore the overlap of strange quark and NSI contributions that enter the NCE scattering cross section of Eq.~\ref{dsdQ}, a two parameter combined analysis is performed. By simultaneously varying the strange quark $g_A^{\mathrm{s}}$ and  NSI $\epsilon_{\mu e}^{uV}$ (setting $\epsilon_{\mu e}^{dV} =0$) parameters, the minimisation of the $\chi^2(\epsilon_{\mu e}^{uV}, \Delta s)$ yields the contours in the parameter space $(\epsilon_{\mu e}^{uV}-\Delta s)$ shown in Fig.~\ref{fig.overlap}, at 68\%, 90\% and 99\% C.L. These results indicate strongly that current neutrino-nucleon experiments are favourable facilities to provide new insights and to put severe bounds on non-standard interaction parameters.

\section{Summary and conclusions}
In the present work, focusing on the NCE neutrino-nucleon scattering, potential corrections to the SM cross sections that arise from strange quark contributions and non-standard neutrino-nucleon interactions are comprehensively investigated. In this context, the possibility of probing the relevant model parameters is explored. Furthermore, special effort has been devoted towards exploring the overlap of possible contributions due to strange quarks and NSI. The study involves reliable calculations of the differential and total neutrino-nucleon cross sections by taking into account important nuclear effects such as the Pauli blocking. Within this framework, the NSI contributions originate from the respective nucleon form factors and adopt dipole momentum dependence, while the corresponding cross sections are rather sensitive to the magnitude of the NSI. The latter have a significant impact on the expected number of NCE neutrino-nucleon events and lead to an enhancement of the rate, which may be detectable by the relevant experiments (e.g. MiniBooNE), even for small values of the flavour changing parameters. It is furthermore shown that possible measurements of the ratio of the NSI cross sections for the $\nu p$ process over the $\nu p$ one offer a unique research path to probe NSI.  

We stress, however, that the above results refer to forward
NSI scattering and thus they do not reproduce accurately
the reported MiniBooNE anomaly where an isotropic excess
of events was found coming either from electrons or from
converted photons. In addition, the NSI contribution would
be small if the ``standard'' value of possible NSI contributions
is chosen. Moreover, in this case, the recoiling protons
within the mineral oil are likely to have velocity below the
Cherenkov threshold and therefore cannot reproduce the
Cherenkov ring. On the other hand, the presence of potential
nonstandard neutrino-nucleon events may be compatible
with the LSND anomaly which did not rely on Cherenkov
radiation.

\section*{Competing Interests}
The authors declare that there are no competing interests
regarding the publication of this paper.

\section*{Acknowledgements}
 The authors are grateful to Prof. S.N. Gninenko for stimulating discussions.

\section*{References}
%




%

\end{document}